\documentclass[pdflatex,sn-mathphys-num]{sn-jnl}

\usepackage{graphicx}%
\usepackage{multirow}%
\usepackage{amsmath,amssymb,amsfonts}%
\usepackage{amsthm}%
\usepackage{mathrsfs}%
\usepackage[title]{appendix}%
\usepackage{xcolor}%
\usepackage{textcomp}%
\usepackage{manyfoot}%
\usepackage{booktabs}%
\usepackage{algorithm}%
\usepackage{algorithmicx}%
\usepackage{algpseudocode}%
\usepackage{listings}%
\usepackage{lmodern}
\usepackage{anyfontsize}

\newcommand{\GG}[1]{\textcolor{blue}{#1}}

\begin{document}

\title[Recent trends in socio-epidemic modelling]{Recent trends in socio-epidemic modelling: behaviours and their determinants}


\author[1]{\fnm{Daniele} \sur{Proverbio}}\email{daniele.proverbio@unitn.it}

\author[1]{\fnm{Riccardo} \sur{Tessarin}}\email{ratessarin@gmail.com}

\author*[1]{\fnm{Giulia} \sur{Giordano}}\email{giulia.giordano@unitn.it}

\affil[1]{\orgdiv{Department of Industrial Engineering}, \orgname{University of Trento}, \orgaddress{\street{via Sommarive 9}, \city{Trento}, \postcode{38123}, \country{Italy}}}




\abstract{The spreading dynamics of infectious diseases is influenced by individual behaviours, which are in turn affected by the level of awareness about the epidemic. Modelling the co-evolution of disease transmission and behavioural changes within a population enables better understanding, prediction and control of epidemics.
\GG{Here, our primary goal is to provide an overview} of the most popular modelling approaches, ranging from compartmental mean-field to agent-based models, \GG{with a particular focus on how behavioural factors are incorporated into epidemic dynamics}.
\GG{We classify modelling approaches based on the fundamental conceptual distinction between models of behaviours and models of behavioural determinants (such as awareness, beliefs, opinions, or trust); in particular, we observe that}
most studies model and interpret the variables related to individual responses either as behaviours or as determinants, with the implicit assumption that they correlate linearly. 
\GG{Based on preliminary empirical observations, we then challenge this assumption} by analysing a recent dataset about time series of social indicators, collected during the COVID-19 pandemic. We examine the case study of Italian regions and we discover that behavioural responses are poorly explained by awareness, beliefs or trust, thereby calling for a careful interpretation of the modelling assumptions and for the development of further models, which fully account for the inherent complexity of individual responses and human behaviours. }



\maketitle

\section{Introduction}\label{sec1}

Modelling epidemic dynamics is fundamental to gain insight into the modes of disease propagation \cite{anderson1991infectious,brauer2012mathematical,Breda2012,Diekmann2000,edelstein2005,Hernandez_Vargas_2022,martcheva2015introduction,pastor2015epidemic}; it enables a better understanding and analysis of the epidemic evolution, the generation of numerical what-if scenarios and predictions \cite{Giordano_2020,Giordano2021,Kemp_2021,Krueger2022,proverbio2024early,proverbio2024data},
as well as the design of mitigation strategies and control interventions \cite{alamo2021data,Donofrio2023,Köhler2020,Lenhart2007,Morris2021,nowzari2016analysis,Proverbio_2021,Sélley2015,Zaric2002}.
Seminal works describing disease propagation within homogeneous populations \cite{kermack1927contribution} gave rise to the class of compartmental mean-field epidemiological models, which have been successfully used for almost a century to capture the spread of infectious diseases; they have been tailored to different diseases and their peculiarities \cite{Brauer2008,Gumel2004,Giordano_2020,Li2020a,PanPox2022} and generalised by including arbitrary numbers of compartments representing different stages of the disease \cite{Arino2007,Blanchini2021,Blanchini2023,CalaCampana2024}.
The mathematical modelling of epidemics has grown to embrace different types of models, including networked multi-patch and meta-population models \cite{bichara2018multi,boguna2013nature,Grenfell1997} that account for the heterogeneity of the epidemic evolution in interconnected geographical areas \cite{Aalto2025,Bertuzzo2020,DellaRossa2020,Gatto2020,Rowthorn2009} or for different age classes \cite{DelValle2013,Hethcote2000,Martcheva2020,Safi2013}.

Recent works have further extended their modelling scope to include additional aspects of the dynamics of human diseases, such as heterogeneity in the patterns of networked social interactions \cite{mei2017dynamics, pastor2015epidemic} or socio-economic aspects \cite{burzynski2021covid, iwasa2023waves}.
As recently shown by the COVID-19 pandemic, societies may be differently impacted by pandemics depending on individual behaviours and compliance with mitigation strategies \cite{donges2022interplay, makki2020compliance}. For instance, people can decide to change their behaviours in response to the perceived risk of being infected \cite{veltri2024assessing}, \emph{e.g.}, by avoiding certain contacts \cite{zhang2020changes}, changing commuting habits \cite{Proverbio_2021} or using personal protective equipment \cite{li2020mask}. Similarly, awareness of the epidemic situation, trust in news outlets and opinion dynamics among peers and on social media can determine the effectiveness of non-pharmaceutical and pharmaceutical interventions \cite{kleitman2021comply,viswanadham2023behavioral}. Therefore, especially for diseases mediated by person-to-person interactions, including models from the social sciences is precious to enhance epidemic models \cite{Bauch_2012_overview, bavel2020using, bedson2021review, funk2015nine}.

Behavioural aspects can be modelled in various ways, depending on the desired level of details, the model scope (in terms of generated understanding, analytical tractability or precision of the predictions) and the availability of data. Different models can be developed to pursue different objectives, \emph{e.g.} capturing the essence of the phenomenon, or faithfully reproducing observed patterns in real-world dynamics. Over the years, numerous modelling approaches have thus been developed, including, \emph{e.g.}, augmenting compartment models with implicit or explicit behavioural dynamics \cite{fenichel2011adaptive, proverbio2024data}, employing networks and multi-layer networks \cite{frieswijk2022mean, peng2021multilayer}, up to using game-theoretic approaches \cite{ye2021game} or agent-based models \cite{palomo2022agent}. 

It is important to consider that models may differ in the interpretation associated with the behavioural terms, which in turn affects the implementation of control strategies and the choice of data for model calibration. In fact, one may either model directly the behavioural responses (such as mask-wearing actions, or behaviour changes, or adherence to vaccination campaigns), or instead try to capture one of the numerous determinants of behaviour, such as awareness, risk attitudes, opinions, trust or beliefs \cite{Funk2009,zino2021analysis, anderson2019recent}. These, in turn, are typically assumed to translate into behaviours, often by assuming a linear relationship. This assumption supports the possibility to leverage results from opinion dynamics theories \cite{xia2011opinion, das2014modelling, bhowmick2020influence,Zino2020},
as well as other models from the social sciences, to mine data from sources such as sentiment databases on social media \cite{pellert2020dashboard, zhu2020analysis} and to use them as proxies of behaviours. Nonetheless, it is still debated whether such linearity assumption accurately captures the influence of behaviours during epidemics \cite{huys2010nonlinear, reitenbach2024coupled}.

To cast light onto the development of models for the co-evolution of epidemics and behavioural responses, this article offers an overview of the most recent trends in the literature; it complements previous reviews, such as those by \citet{Funk_2010, verelst2016behavioural, Wang_2015_review}, by including more recent approaches and perspectives. After surveying a few seminal works from the past decades, we specifically focus on contributions developed after the COVID-19 pandemic, which constituted a significant divide for the epidemiological community. Previous studies \cite{gozzi2024comparative} have segmented the literature into two main modelling classes: Data-Driven behavioural models, which integrate real-world data on behavioural change, and Analytical behavioural models, based on theoretical frameworks. Instead, we offer a classification based on the employed modelling approaches (ranging from compartmental models based on ordinary differential equations to agent-based models), further discussing whether the model gives a behavioural interpretation or focuses on determinants of behaviours, such as awareness or opinions. We discuss the scope of each modelling class and we offer a curated list of key examples from the literature.

In addition, by analysing a recent dataset of questionnaire responses about COVID-19 in Italy, we provide empirical evidence to support the use of distinct models to incorporate behaviours, beliefs and awareness into models of disease dynamics. In fact, these various nuances of behavioural responses do not perfectly correlate and should thus require dedicated functional forms and data integration. 

By systematically classifying modelling approaches, along with their scope and interpretations, this article provides researchers with a set of tools to inform their modelling choices, so as to foster the development of new models to understand and predict the co-evolution of diseases and behavioural responses. 



\section{Overview of behavioural-epidemiological models}

This section reviews the most recent approaches aimed at integrating behavioural responses into epidemiological mathematical models.

We first discuss compartmental mean-field models, such as the Susceptible-Infectious-Removed (SIR) model \cite{kermack1927contribution,brauer2012mathematical,edelstein2005}, which can implicitly incorporate behavioural responses into time-varying parameters or functional expressions, or explicitly include them in the model as additional state variables. These models enable easier analytical treatment and fitting to time-series data; however, due to the underlying homogeneity assumption, they may not capture important details on the complexity of real-world dynamics. We then consider multilayer models, which often explicitly consider the evolution of opinion or awareness dynamics to capture reactive behavioural changes depending on the choices of neighbours and on information spreading. We further survey network models of varying complexity, which introduce coupled dynamics for disease propagation and for social aspects, thus increasing the level of complexity and enabling the observation of non-intuitive patterns. Then, we consider game-theoretic models, which explain behavioural choices, and their co-evolution with diseases, through rational responses that drive decisions based on multi-pronged factors.  Finally, we describe agent-based models, which aim at capturing interactions among individuals within a society as faithfully as possible, albeit at the price of requiring intense computational resources and vast databases.

For all modelling approaches, one may interpret non-epidemic terms as ``behaviours'', thus focusing on what individuals actually \textit{do}, or as ``determinants of behaviours'', such as awareness, opinion, beliefs, risk attitudes -- in this case, the attention is on ``what drives individuals to do what they do''. The first interpretation is more faithful to a behaviouristic paradigm, \textit{i.e.,} the idea that the model's aim is to capture \textit{how} phenomena evolve; the second interpretation aims at exploring \textit{why} phenomena evolve. In principle, the advantage of using the behavioural perspective is to develop more precise predictions, while determinants may be more suitable to inform control strategies as they refer to more fundamental aspects that can be suitably influenced. Using determinants as proxies of behaviours also offers a practical advantage, because determinants can be extracted from sources such as sentiment analysis on social media, while population-wide behaviours can only be captured \textit{a posteriori} via surveys or qualitative investigations.

Throughout this section, while surveying various modelling approaches, we will highlight which ``behavioural'' interpretation is used by each work. In the next section, we will provide empirical evidence as to whether determinants of behaviours can be used as meaningful proxies of behavioural responses in model development.

\subsection{Implicit mean-field models}\label{ref:implicitcompartmental}
\begin{figure}[ht]
    \centering
    \includegraphics[width=0.9\linewidth]{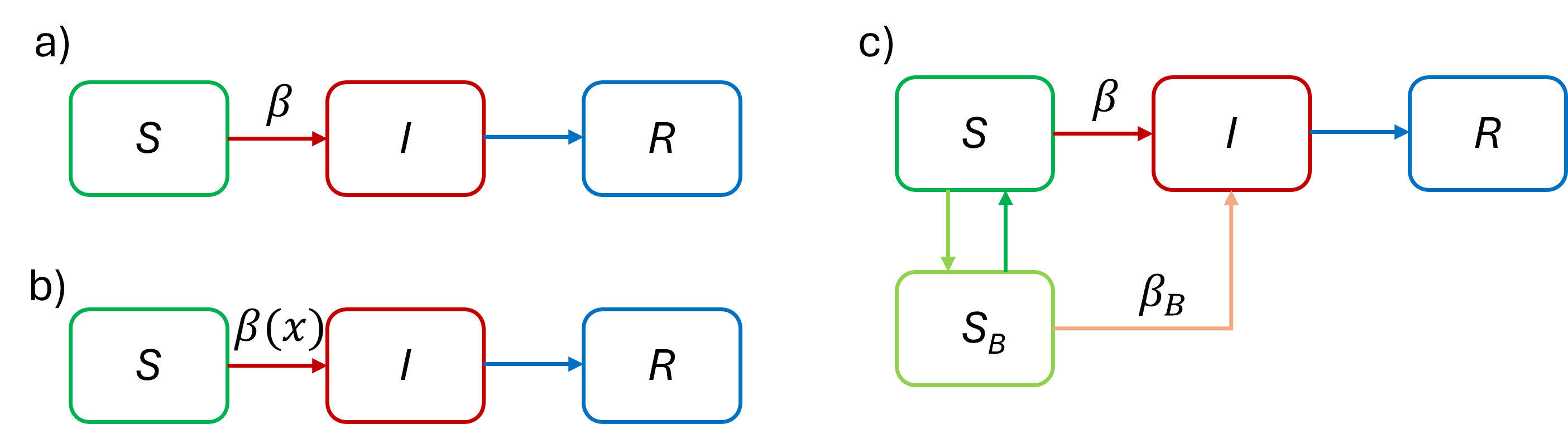}
    \caption{\small Flow diagrams of compartmental models; the influence of $I$ on the infection flow, and parameters other than the contagion rate $\beta$, are not shown for the sake of clarity. (a) The classical SIR model in \cite{kermack1927contribution}. (b) Implicit  modelling of behaviours by turning $\beta$ into a function of other model variables, generically denoted by $x$, as in \cite{capasso1978generalization} where the contagion rate depends on the density of infected $I(t)$, or as in \cite{BuonomoDellaMarca2020} where the contagion rate depends on a suitably defined information index $M(t)$. (c) Explicit modelling of behaviours through an extra compartment $S_B$ representing risk-averse individuals, characterised by a smaller contagion rate $\beta_B < \beta$, as in \cite{gozzi2024comparative}.}
    \label{fig:schemes}
\end{figure}

A first approach to integrate social and epidemiological aspects builds on mean-field compartmental models for epidemics \cite{kermack1927contribution,brauer2012mathematical,edelstein2005}. These models partition the population - which is assumed to be large, well-mixed and homogeneous - into mutually exclusive classes or compartments, each representing a different health stage in relation to the considered disease. We denote by $C$ the set of compartments, which can include, \textit{e.g.}, susceptible, exposed, infectious, infected (either asymptomatic or symptomatic, either detected or undetected), quarantined, vaccinated, recovered individuals. To each compartment, we associate a state variable that represents the density of individuals in that stage.
Transitions between compartments can either be driven by interactions, \textit{e.g.} between susceptible and infectious individuals in the case of contagion (captured by suitable rates of transition from compartment $i \in C$ to compartment $j \in C$ following an interaction between an individual in a healthy stage $i$ and an individual in an infectious stage $c \in C$, where $i \neq j$ but it can possibly be that $j=c$), or depend only on the origin compartment, as in the case of recovery (captured by suitable transition probabilities).
The classical SIR model (Figure~\ref{fig:schemes}a) includes compartments $C = \{S, I, R\}$ that account for Susceptible, Infectious and Removed individuals; the corresponding dynamics of the population densities in each compartment, normalised over a total population numerosity $N$, are given by 
\begin{equation}\label{eq:SIR}
    \begin{cases}
        \dot{S}(t) = - \beta S(t) \, I(t) \\
        \dot{I}(t) = \beta S(t) \, I(t) - \gamma I(t) \\
        \dot{R}(t) = \gamma I(t)
    \end{cases}
\end{equation}
where the parameter $\beta$ is the contagion rate, associated with transitions from $S$ to $I$ driven by interactions between $S$ and $I$, and $\gamma$ is the recovery rate, associated with transitions from $I$ to $R$. The two parameters are coupled by the basic reproduction number \cite{van2017reproduction}, $R_0=\beta/\gamma$, which is the expected number of infections arising from one individual, throughout the entire infectious period, in a population of susceptibles.

\begin{figure}[ht]
    \centering
    \includegraphics[width=0.5\linewidth]{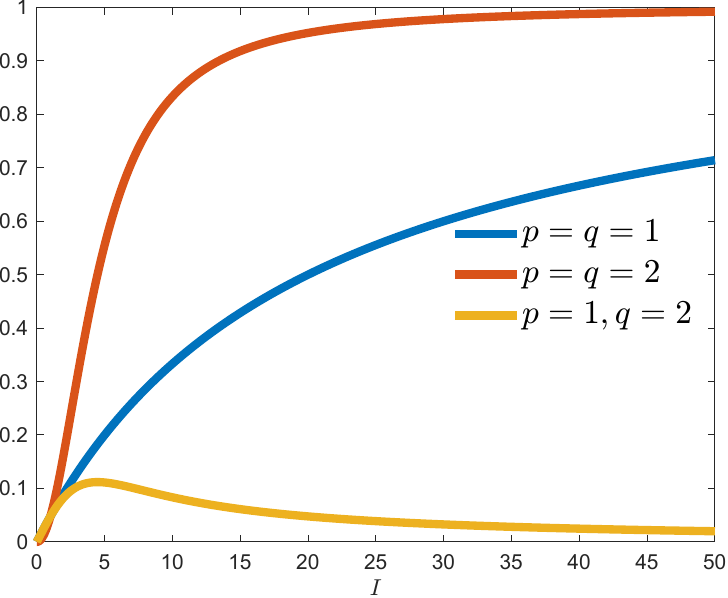}
    \caption{\small Non-linear forces of infection of the Holling type $g(I) = \frac{a I^p}{1+b I^q}$, with $a=b=0.5$, for different choices of $p$ and $q$: $p=q=1$ (blue), $p=q=2$ (red), $p=1$, $q=2$ (yellow).}
    \label{fig:Holling}
\end{figure}

Behavioural responses can be implicitly embedded in these models by making some parameters, such as $\beta$ in \eqref{eq:SIR}, time-varying or state-dependent so as to capture behavioural aspects (see, \textit{e.g.}, Figure~\ref{fig:schemes}b), or equivalently by changing the functional expressions that describe contagion, such as $\beta S(t) I(t)$ in \eqref{eq:SIR}, so as to reflect changes in behaviour that depend on the disease diffusion \cite{fenichel2011adaptive, perra2011towards, alutto2021sir}. For instance, the seminal paper by \citet{capasso1978generalization} considers a SIR model of the form \eqref{eq:SIR} where the parameter $\beta$ becomes dependent on the density of infected, $\beta(I(t))$, or equivalently where, in the contagion term $(\beta I(t)) S(t)$, the linear force of infection $\beta I(t)$ is replaced by the non-linear force of infection $\beta g(I(t))$. The continuous function $g$ is assumed to take nonnegative values, $g(I)\geq0$, with $g(0)=0$, to be upper bounded as $g(I) \leq k$ for all $I$ for some $k>0$, and to have a bounded derivative such that $g'(0)>0$ and $g(I) \leq g'(0)I$ for all positive $I$. A non-linear force of infection takes into account the fact that, although the contagion term can reasonably be a linearly increasing function of the number of infectious individuals when their number is small, saturation effects come into play when $I$ is large, because people are more aware of the epidemic and therefore more cautious. For very large values of $I$, increased awareness and thus precautions can also make the force of infection $g(I)$ decrease as $I$ increases, because individuals drastically reduce their  number of contacts.
The non-linear function $g(I)$, which captures the behavioural feedback on the contagion dynamics, is often chosen as a Holling-type function,
\begin{equation}
    g(I) = \frac{a I^p}{1+b I^q} \qquad \text{with} \qquad p,q>0.
\end{equation}
As shown in Figure~\ref{fig:Holling}, the case $p=q$ corresponds to a saturating function of the Michaelis-Menten type (when $p=q=1$) or of the Hill type (when $p=q$, $p>1$). Conversely, choices such as $p=1$ and $q=2$ model psychological effects due to which $g(I)$ increases as $I$ increases for small values of $I$, while $g(I)$ decreases as $I$ increases for large values of $I$.

Behaviours can be embedded in the contagion parameter also by resorting to saturating functions of \textit{e.g.} the death rates, thus representing the implicit effect of the awareness of the pandemic mortality burden \cite{weitz2020awareness}, or even to more complex functional forms that include perceived caution and safety \cite{usherwood2021model} or social interactions \cite{cabrera2021sir, kolokolnikov2021law, kochanczyk2020dynamics}.

Another approach to implicitly embed behavioural aspects into compartmental models resorts to an integral information index $M(t)$, which depends on the system variables. $M(t)$ captures knowledge about the present epidemic situation along with past information on the spread of the disease, weighted by a time-delayed memory kernel; then, contagion rates and other system parameters depend on $M(t)$ \cite{Buonomo2020,BuonomoDellaMarca2020,Buonomo2021,dOnofrio2007}. This approach has been widely used to study the phenomenon of vaccine hesitancy \cite{Buonomo2018,Buonomo2020,Buonomo2021}, which increases in times of low prevalence due to the reduced perception of danger (thanks to the already acquired herd immunity) and the potential spread of rumours on vaccine side-effects, while the propensity to vaccinate increases during severe outbreaks, due to the increased fear of being infected. 

Implicit behavioural models of the compartmental type can phenomenologically capture the epidemic progression very effectively, by faithfully reproducing the evolution of contagion over time. However, they lack the complexity needed to generate deeper insight into the epidemic process \cite{bavel2020using}. In particular, implicit models cannot distinguish between the various aspects that affect parameter $\beta$, which lumps \textit{e.g.} behavioural feedback, biological changes in viral transmissibility (for instance due to new variants), population-wide non-pharmaceutical interventions such as lockdowns, travel bans, physical distancing, contact tracing and quarantine, use of personal protective equipment \cite{anderson1991infectious,Giordano_2020,Hernandez_Vargas_2022,Hethcote2000,Köhler2020,Proverbio_2021}.

\subsection{Explicit mean-field models}\label{ref:explicitcompartmental}

Several recent studies successfully extend mean-field models with additional compartments that explicitly account for behavioural responses, to investigate specific behavioural changes in closed communities. For instance, \citet{gozzi2024comparative} add a $S_B$ compartment to the classical SIR model, to include susceptible individuals who are risk-averse and are therefore characterised by a smaller contagion rate $\beta_B < \beta$ in view of their protective behaviours (Figure~\ref{fig:schemes}c). In a population where non-pharmaceutical interventions are recommended, \citet{Bongarti2023} investigate compliance to the received recommendation by splitting the susceptible compartment into $S_c$ (compliant) and $S_{nc}$ (non-compliant). \citet{di2021adherence} use a more complex model, accounting for age structure and social contacts, to investigate behaviours that are (or are not) compliant to epidemic control measures. \citet{proverbio2024data} propose an extended behavioural-epidemic model, where all compartments associated with stages of the infection (susceptible, infected, recovered) are partitioned according to differences in behaviour (adoption of measures to avoid becoming infected and infecting others, or lack thereof); the flows between compartments are not only due to the disease evolution, but also capture peer pressure and fatigue that lead to changes in behaviours, as well as homogeneous increases in compliance due to the enforcement of population-wide infection control measures. \citet{ryan2024behaviour} propose a similar modelling approach, where behavioural compartments are interpreted through the lens of the socio-psychological Health Belief Model.

Since these models explicitly include the effect of behaviours on the epidemic dynamics, they have the advantage of enabling the development of targeted control strategies, as well as the use of empirical data for the estimation of parameter values and for model verification. Explicit compartmental behavioural models also allow us to predict the evolution of the epidemic \cite{gozzi2024comparative} and interpret observed data \cite{proverbio2024data}. The main limitation of this model class (in addition to the ideal assumptions of well-mixing and of homogeneity, which characterise all mean-field models) is that several behavioural aspects are lumped together. Network models, which we will discuss later, aim to overcome these limitations.

Explicit compartmental models often rely on determinants of behaviours. For instance, fear can be included by considering additional quarantine compartments for self-protecting individuals \cite{maji2021impact}, while information and misinformation spreading can be captured by models that include apt transitions between informed or misinformed compartments \cite{bulai2024geometric, Sontag2022}. 
Also, several models investigate the impact of awareness on epidemic dynamics. \citet{sahneh2011epidemic} have generalised SIR-like models by including the compartment $A$ of aware, or alert, individuals; such an approach has been further extended, \textit{e.g.}, in \cite{preciado2013convex, sahneh2012existence,zhang2025combined, zino2020assessing, parino2024optimal,Kabir_2019}. Transitions to compartment $A$ are governed by a parameter that represents the effectiveness of awareness campaigns \cite{zino2020assessing}. Overall, the resulting flow diagram is very similar to the one depicted in Figure~\ref{fig:schemes}c, but there is a key difference in the interpretation of the additional compartment, which does not immediately represent \textit{self-protecting} individuals, but \textit{aware} individuals whose self-protecting behaviour is \textit{assumed to linearly correlate} with awareness. Although this seemingly subtle difference may be negligible in the theoretical analysis of such models, it becomes paramount when implementing the devised model-based control strategies and when verifying model predictions by comparing them to empirical data.

Finally, opinion dynamics and social learning (\textit{i.e.}, the idea that opinions that drive behaviours are acquired through interactions among peers \cite{Galesic_2023}) have been explored, \textit{e.g.}, in \cite{Tanaka_2002, Zuo2022, Tyson_2020}, to provide a different interpretation of the drivers of behavioural change and to connect epidemiological models with the flourishing field of opinion dynamics \cite{degroot1974reaching, devia2022framework}. These works include opinion dynamics within the $S$ compartment by partitioning susceptibles depending on their different attitudes toward prophylactic behaviours; a lower probability of infection is associated with being more cautious.

In most models, awareness is assumed to decay due to fatigue or fading memory \cite{mielke2024memory}, yielding transitions to the ``unaware'' $S$ compartment \cite{Wang_2015,Tyson_2020, Tanaka_2002, Zuo2022}. For both awareness and opinion dynamics, the inclusion of additional compartments often represents a first degree of approximation; more complex models, which employ additional variables to account for such behaviour determinants in multi-layer models, are discussed in Section~\ref{sec:att_net}. Moreover, the assumption of a linear relation between behaviour determinants and behaviours can be questioned, as discussed in detail in Section~\ref{sec:corr}.


\subsection{Multilayer models}
\label{sec:att_net}

Explicit mean-field models consider a single variable (a vector of population densities) and a partition of the population into compartments that account both for health status and for behavioural aspects. Conversely, multilayer models couple epidemic and behavioural dynamics by associating each individual $i$ with two variables, $x_i(t)$ representing the health status and $y_i(t)$ representing opinion or behavioural aspects \cite{Sahneh2013}. When $y_i$ represents awareness, it can take binary values, corresponding to aware versus unaware states (for instance, aware individuals can be associated with $y_i=1$ and unaware individuals with $y_i=0$), which in turn elicit protective versus non-protective behaviours, resulting in different infection probabilities \cite{granell2013dynamical, Granell_2014}; alternatively, $y_i(t)$ can take continuous values, representing levels of opinions or beliefs.

The evolution of infectious disease spread and the evolution of opinions, or behaviours, occur therefore in parallel, over two different layers: the physical layer of contagion and the abstract layer of information exchange.
The model thus embeds two co-evolutions, represented as two layers of dynamic interactions among the same individuals, which are associated with nodes in a \textit{network representation} \cite{nunner2021model, nunner2022health, teslya2022effect}, as discussed later in Section~\ref{sec:network}.

\begin{figure}[ht]
    \centering
    \includegraphics[width=0.4\linewidth]{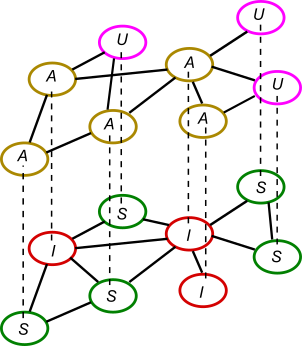}
    \caption{\small Representation of a multilayer epidemic-behavioural structure, as in \cite{granell2013dynamical}. The network, where the nodes are associated with individuals in a population, has two coupled layers: the upper layer describes awareness dynamics (aware, $A$, in gold; unaware, $U$, in purple) and the links capture social influence and communication channels spreading information, while the lower layer describes contagion dynamics (susceptible, $S$, in green; infected, $I$, in red) and the links capture physical contacts potentially spreading the infection. Nodes connected by inter-layer (dashed) connections represent the same individual. The mechanisms spreading contagion and awareness are distinct and occur over two different network topologies; however, both influence the overall co-evolution, because information about the epidemic affects awareness and awareness leads to cautious behaviours that reduce disease transmission.}
    \label{fig:schemes_UAU}
\end{figure}

The base model for the co-evolution of epidemics and awareness is the so-called Unaware-Aware-Unaware model (UAU), often coupled with SIS or SIR dynamics to yield UAU-SIS or UAU-SIR models \cite{granell2013dynamical, Granell_2014}, see Figure~\ref{fig:schemes_UAU}, where $x_i(t) \in \{S, I, R\}$ and $y_i(t) \in \{U, A\}$. Individuals can become aware either after being infected (so, $y_i$ evolves from $U$ to $A$ when $x_i$ evolves from $S$ to $I$), or due to contacts with other aware people. The spread of opinions is modelled through ``epidemic-like'' contagion over the social layer, possibly associated with contacts on a social network, or ``broadcast'' influences from media sources \cite{Misra_2011,Funk2009, guo2022transmission, zhao2021impact, ma2022coupled, xu2023coupled,Collinson2014}. The UAU-SIR model can effectively reproduce emerging dynamics and explosive phenomena, sometimes observed in the real world; it has also provided useful insight in identifying the best social responses to epidemic outbreaks \cite{verelst2016behavioural, viswanadham2023behavioral, Wang_2015}. However, it is unable to capture multi-faceted behavioural responses, which may depend on multiple other factors in addition to awareness \cite{kleitman2021comply}. In fact, identifying to which degree awareness and behavioural responses correlate, hence informing the adoption or refinement of multilayer models based on awareness, is one of the main goals of Section~\ref{sec:corr}.

Another popular socio-epidemic multilayer approach involves opinion dynamics. Opinion dynamics is a rich field in social mathematics, aimed at studying the formation and evolution of opinions in social networks. The most common modelling approach, which often enables analytical results, involves linear averaging dynamics of social influence on individual opinions \cite{anderson2019recent,degroot1974reaching,xia2011opinion}. As in the case of awareness, the additional variable $y_i(t)$, which now represents opinions, is considered within an epidemic model, to capture a second layer of social interactions in addition to the physical layer of contagion. In this case, $y_i(t)$ typically takes continuous values, and evolves according to opinion dynamics models \cite{devia2022framework, Wang_2019}, possibly including influences from epidemic dynamics or peer-pressure from neighbours \cite{anderson2019recent}. Individuals update their opinion and as a consequence, ideally in a direct and linear fashion (whose adherence to reality will be discussed in Section~\ref{sec:corr}), they adjust their behaviour in relation to the epidemic \cite{bhowmick2020influence}. 

How can the model capture the mechanism through which the opinion evolution on one layer affects the spread of contagion on the other layer?
A straightforward coupling corresponds to multiplying, for each individual $i$, the contagion rate $\beta$ of a SIR model by a factor $(1 - \sigma y_i(t))$, with $\sigma \in [0,1]$, to model the effect of self-protection, due to changes in opinions \cite{xuan2020network, bhowmick2022analysis}; individuals in favour of adopting protective behaviours have larger values of $y_i$. This interpretation justifies a time-varying $\beta$, and thus enables model fitting to data \cite{teslya2022effect}, as well as flexible extensions that consider cooperative or polarising opinion dynamics, amenable to analytical results \cite{zanella2023kinetic, she2023epidemics, she2022networked, tunccgencc2021social}. However, as for awareness-based models,  there are limitations in properly capturing the complexity and non-linearity of behaviours, as well as the interplay between behaviours and their determinants \cite{huys2010nonlinear}. The assumptions that behavioural responses are the result of linear opinion dynamics, and that they are directly proportional to opinions, may be too restrictive to enable the prediction and interpretation of actual epidemic patterns.

Traditionally, multilayer models have mostly focused on proxies and determinants of behaviours, coupled with epidemic dynamics. Nonetheless, recent models adopting a multilayer approach have explicitly incorporated behaviours, together with information, awareness or opinion spreading, to build richer and heterogeneous models \cite{xie2024coupled, zuo2021new, wan2022multilayer, zheng2018multiple} (which somehow overlap with the game-theoretic and agent-based approaches described later, at least in their scope). These models account for behaviours explicitly, but often have the downside of being no longer amenable for analytical derivations.

\subsection{Network models}\label{sec:network}

Models that capture the physical network of interactions among individuals, groups, regions or states -- such as multi-patch and meta-population models \cite{Aalto2025,Bertuzzo2020,bichara2018multi,boguna2013nature,DellaRossa2020,Gatto2020,Grenfell1997,Rowthorn2009} -- can help enhance the prediction of epidemic spreading over spatial scales, and account for the heterogeneity of populations and of contact patterns (thus going beyond the well-mixing approximation in mean-field compartmental models).
A network of contacts is represented mathematically by a graph $\mathcal{G}(t) = (\mathcal{N}, \mathcal{M}(t))$, where $\mathcal{N}=\{1, \dots, K \}$ is a set of nodes (representing individuals, or groups, or spatial regions), and $\mathcal{M}(t) \subseteq \mathcal{N} \times \mathcal{N}$ is a (potentially time-varying) set of undirected links, where link $\{n,m\} \in \mathcal{M}(t)$ captures the interaction between nodes $n$ and $m$ at time $t$.

According to the interpretation given to nodes and links, network models can embed various types of epidemic and social dynamics.

If a node $n$ is associated with an individual \textit{within} a compartment, then it is associated with a health state $x_n(t) \in C$ that evolves over time according to the transitions captured by a compartmental model, such as those described in Sections~\ref{ref:implicitcompartmental} and \ref{ref:explicitcompartmental}; the main difference with traditional compartment models lies in the interpretation of parameters, which are probabilities (of contagion, recovery, \textit{etc.}) for a single individual, instead of rates for a whole population. For instance, in a discrete-time networked system, the probability of node $n$ becoming infected, $\mathbb{P}[x_n(t+1) = I | x_n(t)=S]$, depends on the actual number of infectious neighbours in the contact network described by $\mathcal{M}$. This formalism allows one to account for heterogeneity in social contacts (differently from the mean-field compartmental formalism, where each individual is assumed to be equally likely to interact with every other individual in the population) and for stochasticity at the individual level. Typically, networks $\mathcal{G}(t)$ and their dynamics are generated through Markov chain processes \cite{zino2021analysis, nowzari2016analysis, brett2017anticipating,Hernandez_Vargas_2022}, which enable analytical derivations to some degree: instead of studying the time evolution of population densities, one studies the time evolution of infection probabilities. Endowing each individual node $n$ with a second variable $y_n$ that accounts for behavioural states, or for their determinants, leads to the multilayer formalism discussed above \cite{Silva2019, Zuo_2021, peng2021multilayer, Turker_2023}, and exemplified in Figure~\ref{fig:schemes_UAU}.

\begin{figure}[t]
    \centering
    \includegraphics[width=0.5\linewidth]{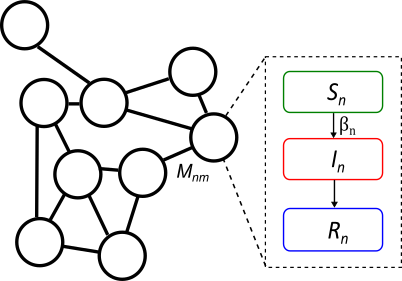}
    \caption{\small A network interconnecting regions, where the epidemic dynamics at each node $n$ is described by a SIR-like model with node-specific parameters; the individual node dynamics are coupled by the links, whose strength is captured by the entries of matrix $M$, where $M_{nm}$ quantifies population mobility and interactions across regions $n$ and $m$ as in \eqref{eq:SIRpatch2}-\eqref{eq:beta}.}
    \label{fig:schemes_net}
\end{figure}

Instead, if nodes represent groups within a society, regions within a country or countries in the world, the network captures dynamic interactions among clusters of people, while links represent social connections among groups or spatial connections among regions or countries (\textit{e.g.}, due to travel patterns). As such, the internal epidemic dynamics at the various nodes are typically captured by SIR-like compartmental models that are interconnected by population mobility and cross-interactions that
couple contagion dynamics at a larger scale \cite{Aalto2025,Bertuzzo2020,boguna2013nature,DellaRossa2020,Gatto2020}; see, \textit{e.g.}, Figure~\ref{fig:schemes_net}. For instance, considering $K$ regions associated with the graph nodes, the evolution of a multi-patch SIR model is captured by
\begin{equation}\label{eq:SIRpatch1}
    \begin{cases}
        \dot{S}_n(t) = - S_n(t) \sum_{m=1}^K \beta_{nm} I_m(t) \\
        \dot{I}_n(t) = - \gamma_n I_n(t) + S_n(t) \sum_{m=1}^K \beta_{nm} I_m(t) \\
        \dot{R}_n(t) = \gamma_n I_n(t)
    \end{cases}
\end{equation}
for all $n=1,\dots,K$, where $(S_n,I_n,R_n)$ represent the fractions of susceptible, infected and recovered individuals in the $n$-th region, $\gamma_n$ is the recovery rate in the $n$-th region and $\beta_{nm}$ is the contagion rate due to contacts between a susceptible individual in the $n$-th region and an infected individual from the $m$-th region; $\beta_{nm}=0$ if no link connects regions $n$ and $m$.

Alternatively, by considering $I(t)=(I_1(t),\dots,I_K(t))$, one can write
\begin{equation}\label{eq:SIRpatch2}
    \begin{cases}
        \dot{S}_n(t) = - S_n(t) \beta_{n}(I(t)) \\
        \dot{I}_n(t) = - \gamma_n I_n(t) + S_n(t) \beta_{n}(I(t)) \\
        \dot{R}_n(t) = \gamma_n I_n(t)
    \end{cases}
\end{equation}
where the contagion parameter $\beta_n$ in the $n$-th region is modelled as a weighted sum over infectious elements that are in contact with node $n$, yielding:
\begin{equation}
    \beta_n(I(t)) = \hat{\beta}_n \sum_{m=1}^K M_{nm} I_m(t),
    \label{eq:beta}
\end{equation}
where $\hat{\beta}_n$ is the contagion rate for the country, or region, in isolation, while $M_{nm}$ denotes the strength of contacts between nodes $n$ and $m$.
Note that \eqref{eq:SIRpatch2}-\eqref{eq:beta} can be cast in the form \eqref{eq:SIRpatch1} by setting $\beta_{nm}=\hat \beta_n M_{nm}$.
The contact matrix $M=(M_{nm})_{1 \leq n,m \leq K}$ is not normalised and already accounts for the proportion of mobile infectious cases; alternatively, one may consider a matrix $M$ with entries that are either $0$ or $1$, and thus only account for the contact pattern, and multiply each term in the sum in \eqref{eq:beta} by an additional parameter that accounts for the effective contacts or mobility of individuals between nodes, as done in \cite{zhou2020effects}.

In general, network models have been developed to capture changes in behaviours with regard to social mixing and mobility habits \cite{Harris_2022, Oh2021}, often coupling computational insight, mostly generated with numerical simulations, with empirical data, collected via mobility trackers, travel statistics or traffic of mobile phones \cite{oliver2020mobile, alessandretti2022human, hu2021human, zhou2020effects, lai2020effect, wu2020nowcasting, chang2021mobility, Tizzoni2014}, or even from mobility reports such as those from Google \cite{google2020, prestige2023estimating}. Network models have also been employed to capture changes in working and social habits within cities and countries \cite{burzynski2021covid, zino2021analysis}. Multi-patch network models reproduce the observed contact patterns among groups or regions, but hardly capture changes in behaviours at the individual level; therefore, they are not particularly useful to design targeted strategies to inform single individuals about, \textit{e.g.}, prophylactic behaviours such as mask wearing or personal hygiene.

Alternatively, network models can be developed and informed using data and patterns extracted from social networks \cite{pawelek2014modeling, molaei2019predicting}, which provide useful sources of information about opinions, awareness, sentiment and attitudes about an epidemic, with fine-grained heterogeneity and spatial resolution \cite{alvarez2017epidemic, zhu2020analysis, carballosa2021incorporating, zhang2020new}. These models try to capture the co-evolution of determinants of behaviours and epidemics, but rely mostly on data-driven approaches and less on theoretical analysis. Moreover, by relying on proxy data, mostly extracted from social networks and only weakly correlating with actual changes in behaviour, they may miss important non-linear patterns and they often require validation via formal analysis and empirical surveys.

\subsection{Game-theoretic models}

Game-theoretic models have been used to explore strategic interactions between individuals, capturing the complexity and non-linearity of human decision-making processes to model how individuals and populations adjust their behaviour during an epidemic. The framework assumes that each individual $n$ within a population can choose between a set of actions; the choice is captured by a variable $z_n(t)$. If there are only two viable actions, for instance accepting or refusing vaccination, $z_n(t) \in \{0,1 \}$ and each individual chooses the value of $z_n$ after a rational comparison of two payoff matrices, associated with the two different actions, which can depend on many epidemic and social factors, such as the current herd immunity level, the number of infected individuals, the personal risk attitude, social pressure, or tendencies to homophily \cite{di2015quantifying, reitenbach2023coupled, nabi2024analyzing, Smaldino_2021}. The framework is extremely flexible and can accommodate several drivers of behavioural change, with increasing levels of complexity.

Game-theoretic models have been first applied to the study of voluntary vaccination campaigns. The seminal works \cite{auld2003choices, bauch2004vaccination}, which feature basic payoff matrices and decision mechanisms, have been later extended with additional features such as imperfect vaccination, different time-scales and non-linearities in human decision-making \cite{chen2019imperfect,della2024geometric,Hota2019,fu2011imitation}. Nowadays, game-theoretic models are no longer confined to vaccination campaigns, but have covered several behavioural aspects, including social distancing, mask wearing, or response to recommended non-pharmaceutical interventions \cite{kabir2020evolutionary, brune2020evolutionary, nagurney2021game}, as well as the interplay with awareness, beliefs and opinions \cite{ye2021game, de2024game, reluga2010game, Bauch_2012_game, wang2020}. The payoff functions have also been extended to include fatigue and memory, and account for social influence and relations between behaviours and their determinants \cite{maitra2023sis, elokda2021dynamic, satapathi2023coupled, ye2021game}.
The game-theoretic approach has also been used to design optimal intervention policies \cite{martins2023epidemic}, and some game-theoretic models have been validated through comparison with empirical data \cite{Gosak_2021_game, wells2020prosocial}, to showcase their effectiveness in capturing changes in behaviours that occur after rational pondering. In addition to payoff matrices, complex game-theoretic models embed epidemic dynamics over networks, where nodes are also equipped with variables $z_n$ to model their decisions. A recent example of implementation is described in \cite{ye2021game}, while surveys about game-theoretic modelling of infectious diseases are in \cite{chang2020game, huang2022game}.

\subsection{Agent-based models}
Agent-based models (ABMs) simulate the progression of a disease by focusing on the behaviour and interaction patterns of autonomous agents, which represent the computational replicas of either individuals or collective entities (such as organisations). This modelling approach reproduces the observed spontaneous interactions between such agents, yielding the emergence of disease spreading \cite{cliff2018investigating, hackl2019epidemic}. However, this approach does not allow for analytical results. Its main goal is to understand how individual behaviours, following a set of rules, influence the overall evolution of the system. Agent-based models naturally embed behaviours in their definition of agents, and have been developed both for generic infectious diseases and for specific epidemics such as COVID-19 or malaria \cite{smith2018agent, perez2009agent, hoertel2020stochastic}. By simply adding appropriate interaction rules in the definition of the computational agents, ABMs can capture the impact of awareness or fear on epidemic spreading, and can account for how sharing opinions among agents affects individual responses and disease propagation \cite{paarporn2015epidemic, maziarz2020agent, yu2024individuals, guo2021modeling, Epstein_2021}. 

Agent-based models are very versatile and can reproduce social and epidemic patterns with high fidelity. For an overview of this field, we refer the reader to recent surveys such as \cite{nianogo2015agent, lorig2021agent, hunter2017taxonomy}.
A key advantage of ABMs is that they provide an intuitive tool to interpret epidemic modelling by focusing on individual perspectives, which helps propose individual-centred policies \cite{niemann2024multilevel,silva2020covid}. Moreover, they require minimal assumptions and allow one to include more complex scenarios and requirements than game-theoretic models.
As for their limitations, ABMs are usually not amenable to formal analysis and mostly rely on massive computational exploration of the parameter space, which, due to the huge computational complexity of the models, may hinder the computation of optimal control strategies. Also, to generate plausible scenarios and identify meaningful parameter values, ABMs require extremely large amounts of detailed and reliable data.
Collecting and incorporating these data can be a challenging task, especially when data about social, behavioural and psychological attitudes and changes are considered. Efforts are made to use direct behavioural data or proxy data but, as of today, these challenges remain a practical limitation of this modelling approach \cite{alamo2021data}.

\subsection{Discussion}

Several modelling approaches have been proposed to integrate epidemic and social dynamics. They vary depending on the adopted framework (from mean-field compartmental models to agent-based models), and in the use and interpretation of their state variables, which may represent behaviours either directly, or through their determinants. We have presented an overview of different modelling approaches, which we have classified into coarse-grained classes. It is worth pointing out that several approaches may coexist, and hybrid models have been successfully developed, \textit{e.g.}, by considering multilayer network models (often called multiplex networks). The essential bibliography with pointers to recent studies is summarised in Table~\ref{tab:summary}.

\begin{table}[b]
\centering\footnotesize 
\begin{tabular}{|c|c|c|}
\hline
Model & Behaviour & Determinant(s)  \\ \hline
Implicit mean-field  & \cite{fenichel2011adaptive, perra2011towards, alutto2021sir, capasso1978generalization, cabrera2021sir, kolokolnikov2021law, kochanczyk2020dynamics}  & \cite{weitz2020awareness, gozzi2024comparative, usherwood2021model} \\ 
Explicit mean field   & \cite{gozzi2024comparative, Bongarti2023, di2021adherence, proverbio2024data,Wang_2015,dOnofrio2007,Buonomo2018,Buonomo2020,BuonomoDellaMarca2020,Buonomo2021}  & \cite{maji2021impact, bulai2024geometric, sahneh2011epidemic, preciado2013convex, sahneh2012existence, zhang2025combined, zino2020assessing, parino2024optimal, Tanaka_2002, Zuo2022, Tyson_2020,Kabir_2019,Sontag2022} \\ 
Multilayer    &  \cite{nunner2021model, nunner2022health, xie2024coupled, zuo2021new, wan2022multilayer, zheng2018multiple} &  \cite{teslya2022effect, granell2013dynamical, Granell_2014,Misra_2011, Funk2009, guo2022transmission, zhao2021impact, ma2022coupled, xu2023coupled, verelst2016behavioural, viswanadham2023behavioral, Wang_2015, anderson2019recent, xia2011opinion, bhowmick2020influence, xuan2020network, bhowmick2022analysis, zanella2023kinetic, she2023epidemics, she2022networked, tunccgencc2021social,Collinson2014}   \\ 
Network & \cite{burzynski2021covid, zino2021analysis,oliver2020mobile, alessandretti2022human, hu2021human, prestige2023estimating, zhou2020effects, lai2020effect, wu2020nowcasting, chang2021mobility}  &  \cite{pawelek2014modeling, molaei2019predicting, alvarez2017epidemic, zhu2020analysis, carballosa2021incorporating, zhang2020new, Silva2019, Zuo_2021, peng2021multilayer, Turker_2023} \\
Game-theoretic       &  \cite{auld2003choices, bauch2004vaccination, chen2019imperfect, della2024geometric, fu2011imitation, kabir2020evolutionary, brune2020evolutionary, nagurney2021game, maitra2023sis, elokda2021dynamic, satapathi2023coupled, martins2023epidemic, Gosak_2021_game, wells2020prosocial, chang2020game, huang2022game}  &  \cite{ye2021game, de2024game, reluga2010game, Bauch_2012_game, wang2020} \\ 
Agent-based  & \cite{smith2018agent, perez2009agent, hoertel2020stochastic, nianogo2015agent, lorig2021agent, hunter2017taxonomy, niemann2024multilevel,silva2020covid} & \cite{paarporn2015epidemic, maziarz2020agent, yu2024individuals, guo2021modeling, Epstein_2021}  \\
\hline
\end{tabular}
\caption{The recent works surveyed in this article, partitioned into the different considered modelling frameworks, are classified depending on whether they consider variables (or data) that can be interpreted as ``behaviours'' or as ``determinants of behaviour''.}
\label{tab:summary}
\end{table}

In our overview, we have seen that considering behaviours or their determinants is mostly a matter of modelling choice, and the linear relationship between determinants and behaviours is typically an implicit assumption in the literature. Such an assumption can be questioned when thinking about non-linearities in human decision-making \cite{huys2010nonlinear,Hota2019}, but it is rarely tested, even via model fitting. In Section~\ref{sec:corr}, we thus challenge this assumption by analysing an empirical dataset, which includes time-series of behaviours and of other indicators collected during the COVID-19 pandemic.

\section{Correlation between behaviours, beliefs and awareness: empirical evidence}
\label{sec:corr}
As we have seen, behavioural terms in epidemiological models are often interpreted through the use of behaviour determinants, such as awareness, opinions or beliefs, and a linear relationship is assumed between behaviours and their determinants; this allows the use of proxy data to calibrate the models. To shed light onto the appropriateness of assuming such a linear relationship, and thus inform the development of future epidemiological-behavioural models, in this section we provide novel empirical results extracted from a recent dataset, created during the COVID-19 pandemic \cite{astley2021global}.

\subsection{The empirical dataset}

The socio-behavioural dataset we consider was developed by the initiative ``Global COVID-19 Trends and Impact Survey'' of the University of Maryland Social Data Science Center, in partnership with Meta (Facebook). The dataset collects answers (more than $100,000$ per day) to questionnaires filled out by Facebook users, and covers most countries in the world throughout a period from 2020 to 2022. All data were anonymised, weighted, normalised, curated, aggregated and made available through API by the University of Maryland, as described in \cite{barkay2020weights}. 
At the time of writing, the dataset represents the most abundant collection of time series available for data mining and analysis, beyond traditional surveys. Despite natural limitations, such as coverage, associated with any web survey and disclosed on the project website (\url{gisumd.github.io/COVID-19-API-Documentation/}), it provides a valuable source to test hypotheses about trends in behaviours, awareness, trust, socio-economic factors, attitudes, and so on. For instance, it has been employed to uncover trends in sudden behaviour changes during the early phases of the COVID-19 pandemic \cite{heino2023attractor}. 

The data are grouped into time-series arrays of indicators, divided by domains such as symptoms, behaviours, economic factors, mental health. For each indicator, the dataset reports counters about the (normalised) number of respondents who answered positively to the associated question. The full list of indicators and their explanation can be found on the project website. Since attitudes towards vaccination were particularly prone to information and misinformation campaigns and were highly politicised, we focus instead on mask wearing behaviours during the COVID-19 pandemic, specifically looking at Italy with a resolution at the regional level. Some regions were associated with a low number of respondents, or a low frequency of reported responses; therefore, we select regions with a sufficient representativeness to allow for statistical analysis: Campania, Emilia-Romagna, Lazio, Lombardia, Piemonte, Puglia, Sicilia, Toscana, Veneto. The considered time period, for which data were consistently available for all indicators, ranges from May 21$^{st}$, 2021 to June 25$^{th}$, 2022.

In addition, we collect incidence (daily cases) data about the time evolution of the COVID-19 pandemic from the Italian Protezione Civile Github repository, \url{https://github.com/pcm-dpc/COVID-19}, during the same time period. Finally, we consider data regarding non-pharmaceutical interventions and their stringency, based on the Oxford COVID-19 Government response tracker and Stringency Index \cite{hale2021global}, which capture the stringency of the policies adopted by local governments, regarding the enforcement of certain behaviours, on a scale from 1 (``No policy'') to 5 (``Required outside-the-home at all time'').

\subsection{Social indicators}
The socio-behavioural dataset is subdivided into indicators. Consistently with the scope of our analysis, we primarily employ the indicator ``mask'' (behavioural indicator), which counts how many respondents (in percentage) declared they wore a mask all the time or most of the time when in public.  

The other selected indicators relate to beliefs and awareness. 

The first belief indicator, $B_1$, relates to people who are worried about catching COVID-19; the second indicator, $B_2$, captures the proportion of people who believe that wearing masks is effective to prevent contagion. 

\begin{figure}[h]
    \centering
    \includegraphics[width=\linewidth]{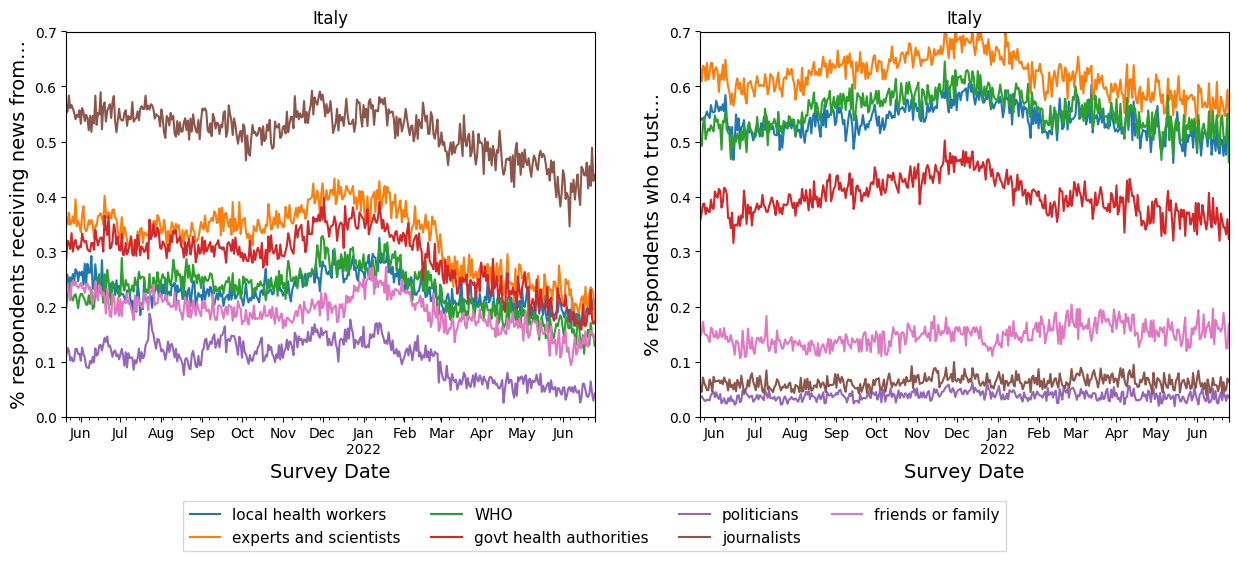}
    \caption{\small Time series for indicators of awareness (a) and trust (b) concerning seven different sources of information.}
    \label{fig:awareness_trust}
\end{figure}

For awareness, we use seven different indicators, $A_1$ to $A_7$, that measure the proportion of respondents who received news from 1) local health workers, clinics, and community organizations; 2) scientists and other health experts; 3) the World Health Organization (WHO); 4) government health authorities or officials; 5) politicians; 6) journalists; 7) friends and family. Figure~\ref{fig:awareness_trust}a shows time-series data for awareness indicators during the considered time period, for Italy as a whole. 

Awareness about a source of information has a different impact on determining behaviours, depending on the trust placed upon that source; for instance, the time evolution of trust in the seven considered news sources in Italy during the considered time period is shown in Figure~\ref{fig:awareness_trust}b. However, we do not explicitly consider trust in our analysis, since i) we did not find models in the literature that explicitly account for this aspect; ii) the functional form expressing how awareness should be weighted by trust is unknown, and \citet{proverbio2024data} have observed that using a simple linear scaling offers no significant advantages; iii) data about trust are sparser than those available for other indicators, and so the statistics are hardly comparable.

\subsection{Analysis and results}

\begin{figure}[h]
    \centering
    \includegraphics[width=\linewidth]{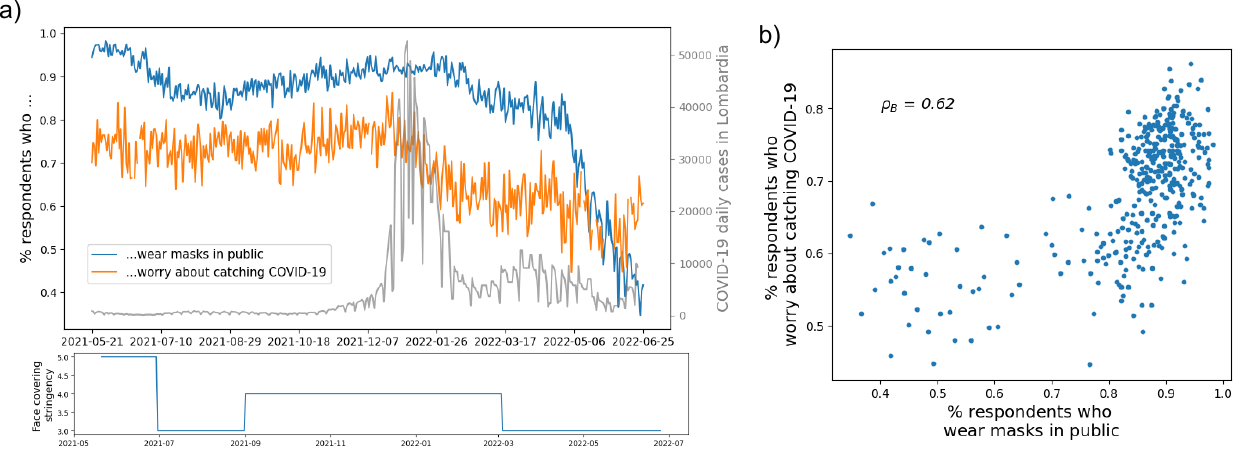}
    \caption{\small Example of data analysis for Lombardia. a) Time series of the indicators of mask wearing in public (behaviour, blue) and worry about catching COVID-19 (belief $B_1$, orange). As a reference, we superimpose the dynamics of incidence data for Lombardia. Below, we also show the evolution of the stringency index for mask wearing enforcement in Italy. b) Correlation between the indicators of mask wearing (behaviour) and worry about catching COVID-19 (belief); the resulting Pearson's coefficient is $\rho = 0.62$.}
    \label{fig:corr}
\end{figure}

For all considered Italian regions, we check whether behaviours linearly correlate with beliefs and awareness. To do so, we calculate the pair-wise Pearson's correlation coefficient $\rho$ of time-series data for various indicators, used as dynamical variables. For instance, Figure~\ref{fig:corr} shows that in Lombardia worrying about catching COVID-19 (belief) correlates with $\rho = 0.62$ with wearing masks in public (behaviour).

\begin{figure}[ht]
    \centering
    \includegraphics[width=0.9\linewidth]{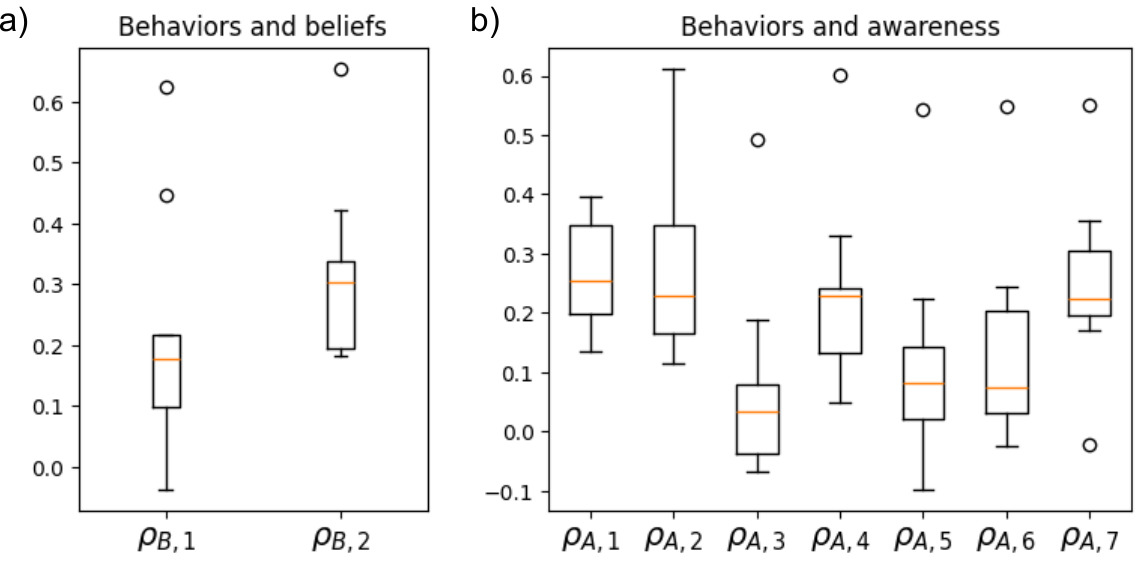}
    \caption{\small Distributions of Pearson correlation coefficients between scores of (a) behaviours and beliefs, $\rho_{B,i}$, and (b) behaviours and awareness, $\rho_{A,j}$. For the belief indicators corresponding to $i=\{1,2 \}$ and the awareness indicators corresponding to $j=\{1 \dots 7 \}$, refer to main text. }
    \label{fig:distributions}
\end{figure}

By estimating correlations for all regions, we extract the distributions of $\rho_{B,i}$, with $i=\{1,2\}$, for behaviour and belief indicators, and $\rho_{A,j}$, with $j=\{1,7\}$, for behaviour and awareness indicators. The results are shown in Figure~\ref{fig:distributions}. Our results are consistent with the hypothesis in the literature \cite{nunner2021model} that behaviours do not perfectly correlate over time with beliefs or awareness. Overall, mask-wearing behaviours correlate weakly with awareness, but not in a perfectly linear fashion ($\rho$ ranges from 0 to about 0.6). The result also heavily depends on the considered region and news source, as indicated by the wide distributions that highlight the lack of homogeneity across the country. For instance, receiving news from health workers and experts better correlates with mask wearing in most regions, while other sources elicit more outliers in the distributions, with lower medians. Figure~\ref{fig:corr}a clearly shows that behaviours can also be influenced by the stringency of the enforced policy, which affects both beliefs and behaviours in a non-trivial fashion.

These results suggests that, on average, using proxy data based on awareness or beliefs may be a reasonable first approximation to calibrate models, but refinements are necessary to better capture socio-epidemic dynamics. Directly employing behavioural data and related dynamics is thus recommended to overcome the limitation due to inaccuracies stemming from non-linear relationships between behaviours and their determinants. Alternatively, modellers are suggested to first test their hypothesis of linear relationship on the region/country of interest, recalling the heterogeneity in the correlations; for instance, another analysis performed on European countries observed higher correlations in northern countries \cite{proverbio2024data}.

\section{Conclusion and perspectives}

Modelling the co-evolution of epidemics and behavioural responses can help better reproduce and predict the spread of infectious diseases, identify targeted communication and intervention strategies, and design measures for epidemic control that are both effective in containing contagion and compliant with civil rights. In addition to the rich literature of results obtained in the past decades, the recent COVID-19 pandemic spurred myriads of new models, employing different approaches to better shed light onto the epidemic evolution in various ways -- from enhancing predictions, to generating new insight via formal analysis, up to informing policy-making.

\GG{This survey provides both an overview and} a classification of such approaches, with a particular emphasis on the interpretation of the variables representing social or individual attitudes. 
In fact, some models directly aim at capturing behavioural responses, while others focus on modelling determinants of behaviours, such as awareness, opinions, fear, beliefs. Explicitly embedding behaviour-related variables and parameters has the advantage of offering higher fidelity with respect to observed patterns and empirical data, while focusing on determinants may be useful if the goal is to inform interventions targeted to individuals, such as information campaigns. Nonetheless, a key question (often implicitly answered positively in the modelling assumptions, without an explicit discussion) is whether behaviours and determinants correlate linearly, and thus enable proportional simplifications of their dynamics.

\GG{Our brief empirical analysis provides a first evidence} that the linearity assumption may be misleading: the analysis of an empirical database collecting individual responses about awareness, behaviours, beliefs and trust, with a focus on Italy, reveals that determinants and behaviours correlate weakly, with significant differences even among regions of the same country. Instead, behaviours have a complex dependence on additional drivers, such as the stringency of the adopted policies.

Our results suggest a critical evaluation of modelling assumptions, as well as a careful interpretation of model results, depending on the nature of the considered variables. They also emphasise the need for more refined models, capturing realistic dependencies between determinants and behaviours: new implicit and explicit mean-field models may be developed by introducing more nuanced functional forms, while other modelling approaches may benefit from our insight by introducing non-linearities in their dynamics.

\GG{The primary contribution of our survey is expository and synthetic, and our preliminary empirical observations represent a first step towards future modelling directions, as well as the validation and interpretation of behavioural-epidemic models.}
Identifying common patterns that relate determinants and behaviours across countries remains a challenge. A thorough analysis of the ``Global COVID-19 Trends and Impact Survey'' dataset may provide more information, along with the development of new datasets and advances in data-mining techniques. Closer multidisciplinary collaboration between modellers and social and behavioural scientists can yield precious insight. Benchmarking different modelling approaches, accounting for their modelling scope and prediction capabilities, will also help validate and identify a set of core models that are more suitable for practical applications. 

Despite the challenges posed by the suggested increase in complexity and the need to validate non-linear dependencies, leveraging our observations to develop the next generation of behavioural-epidemic models will be crucial to unravel emerging patterns, integrate new data sources, improve predictions, and develop more effective control strategies to face present and future pandemics.

\section*{Acknowledgements and declarations}
This work was funded by the European Union through the ERC INSPIRE grant (project number 101076926) and the Next Generation EU, Mission 4, Component 2, PRIN 2022 grant PRIDE (project number 2022LP77J4, CUP E53D23000720006).
Views and opinions expressed are however those of the authors only and do not necessarily reflect those of the European Union, the European Research Council Executive Agency or the European Council. Neither the European Union nor the European Research Council Executive Agency or the European Council can be held responsible for them.

All authors declare that they have no competing interests.

\small
\bibliography{sn-bibliography}

\end{document}